\documentclass[lettersize,journal]{IEEEtran}
\usepackage{amsthm,amssymb,amsmath,amsfonts}
\theoremstyle{plain}

\usepackage[percent]{overpic}
\usepackage{algorithmic}
\usepackage{algorithm}
\usepackage{array}
\usepackage[caption=false,font=normalsize,labelfont=sf,textfont=sf]{subfig}
\usepackage{textcomp}
\usepackage{stfloats}
\usepackage{url}
\usepackage{xcolor}
\usepackage{verbatim}
\usepackage{graphicx}
\usepackage{subfig}
\usepackage{cite}
\usepackage[hidelinks]{hyperref}
\makeatletter
\newcommand*\bigcdot{\mathpalette\bigcdot@{0.7}}
\newcommand*\bigcdot@[2]{\mathbin{\vcenter{\hbox{\scalebox{#2}{$\m@th#1\bullet$}}}}}
\makeatother
\begin{document}

\title{Low-PAPR Joint Channel Estimation and Data Detection  in ZP-OTFS System }

\author{Omid Abbassi Aghda, Mohammad Javad Omidi, Hamid Saeedi-Sourck
        % <-this % stops a space
\thanks{O. A. Aghda and M.J. Omidi are with the Dept. of Electrical and Computer Engineering, Isfahan University of Technology, Isfahan 84156-83111, Iran (e-mail: omidi@iut.ac.ir, o.abasi@ec.iut.ac.ir). M.J. Omidi is also with the Dept. of Electronics and Communication Engineering, KCST, Doha 35003, Kuwait. H. Saeedi-Sourck is with the Dept. of Electrical Engineering, Yazd University, Yazd 89158-18411, Iran. (e-mail: saeedi@yazd.ac.ir).}% <-this % stops a space
}


\newcommand{\mychar}{%
	\begingroup\normalfont
	\includegraphics[height=\fontcharht\font`\B]{picture.png}%
	\endgroup
}
\maketitle

\begin{abstract}
\textbf{Orthogonal Time Frequency Space (OTFS) systems face significant challenges in channel estimation due to high pilot overhead and peak-to-average power ratio (PAPR). To address these issues, we propose a  two-step channel estimation method for Zero-Pad OTFS (ZP-OTFS), a modified OTFS system characterized by multiple zero rows along the delay axis. This  method strategically inserts pilot sequences into the zero bins of the ZP-OTFS system, effectively mitigating overhead and PAPR. Comprehensive simulation results validate the effectiveness of our proposed method, demonstrating its superior performance over traditional embedded pilot estimation in high Signal-to-Noise Ratio (SNR) scenarios. Specifically, our method achieves a lower normalized mean square error (NMSE) and better bit error rates (BER) at high SNRs.}
\end{abstract}

\section{Introduction}
%\IEEEPARstart{R}{eliable}  communication is essential in high-mobility environments for the 6th generation of wireless telecommunication systems. However, in such environments, orthogonal frequency division multiplexing (OFDM)  is not effective due to a high Doppler shift, which results in high inter-carrier interference and degrades its performance \cite{Hadani2017}. To address this issue, a new modulation technique called orthogonal time frequency space (OTFS) was proposed in \cite{Hadani2017}. OTFS modulation multiplexes information symbols into the Delay-Doppler (DD) domain, where each information symbol spreads over the entire time-frequency (TF) domain. As a result, it enables full channel diversity to be exploited. The DD domain is a representation of the channel in terms of its delay and Doppler characteristics, while the TF domain is a representation of the signal in terms of its time and frequency components \cite{Hadani2017}. Despite the advantages of OTFS, data detection and channel estimation remain challenging issues that need to be addressed. 

\IEEEPARstart{R}{eliable} communication in high-mobility environments is crucial for the 6th generation of wireless telecommunication systems. Orthogonal frequency division multiplexing (OFDM) struggles in these environments due to high Doppler shift, leading to significant inter-carrier interference. Orthogonal time frequency space (OTFS), a new modulation technique, addresses this by multiplexing information symbols into the delay-Doppler (DD) domain, allowing full channel diversity exploitation. The DD domain represents the channel's delay and Doppler characteristics, while the time-frequency (TF) domain represents the signal's time and frequency components. However, data detection and channel estimation in OTFS remain challenging \cite{DDCbook}.

Data detection in OTFS systems can be computationally demanding, particularly when using conventional methods such as zero-forcing (ZF) and minimum mean square error (MMSE). To address this challenge, a low-complexity linear MMSE (LMMSE) detection method was proposed \cite{Tiwari2019}. This method takes advantage of the structure of the DD channel matrix to reduce detection complexity. Additionally, the message passing (MP) algorithm has been used for data detection \cite{Raviteja2018a}, offering significantly lower computational complexity compared to MMSE and ZF detectors. The lower computational complexity of the MP detector is due to its ability to leverage the sparsity of the DD channel matrix.  However, despite its lower complexity compared to MMSE and ZF detectors, the MP algorithm still has high complexity. The practicality of the MP detector may also be limited by its dependence on frame size, the number of channel taps in the DD domain, and the constellation size of the quadrature amplitude modulation (QAM) symbols \cite{Raviteja2018a}. In conclusion, further research is needed to develop data detection methods with even lower computational complexity for OTFS systems.
 To further reduce detection complexity, the authors in \cite{Thaj2020} proposed a delay-time (DT) maximal ratio combining (MRC) detector for use in OTFS systems. The MRC detector has significantly less computational complexity compared to MP and LMMSE detectors since its complexity only depends on the frame size and the maximum delay-spread of the channel. Furthermore, the MRC detector improves the signal-to-noise ratio (SNR) of the received signal, resulting in a lower bit error rate (BER). However, in order to use the MRC detector, a modified version of OTFS known as zero-pad OTFS (ZP-OTFS) must be employed. ZP-OTFS was specifically designed for this purpose and involves setting multiple rows along the delay axis to zero. While this zero-padding technique enables the use of the MRC detector, it also introduces high system overhead, which is a significant disadvantage of ZP-OTFS. Overall, the MRC detector offers benefits such as lower complexity and improved SNR, but its use requires ZP-OTFS, which introduces high system overhead.

%This paper proposes a pilot design with a two-step channel estimation algorithm for ZP-OTFS systems. The proposed design addresses the disadvantages of using ZP-OTFS by inserting pilot sequences in zero bins.  Such design enables us to use the MRC detector, offering the benefits of lower computational complexity and improved SNR. The proposed method also utilizes the time domain received sequences to estimate the CSI which leads to reduced estimation complexity.

Accurate estimation of the channel state information (CSI) is vital for wireless communication systems, as the detector’s performance heavily depends on it. The embedded pilot (EP) method, proposed in \cite{Raviteja2019}, is a way to estimate CSI in OTFS. Due to its use of the sparse nature of the DD channel representation and its threshold-based detection of channel taps, it has a low estimation error and low computational cost, respectively. However, the EP scheme has two drawbacks: it suffers from a high peak-to-average power ratio (PAPR) due to the high power of the pilot impulse, and it has high pilot overhead caused by the need to separate the single pilot from information symbols in the DD domain with guard symbols to prevent interference at the receiver.
To address this overhead and take advantage of the sparse representation of the wireless channel in the DD domain, a channel estimation method based on a sparse recovery algorithm was explored in \cite{Zhao2020}. While this method outperforms EP channel estimation, it requires more total power dedicated to pilot sequences, which could be challenging and impose high impractical PAPR to the system.
In contrast, \cite{farhang} proposes a channel estimation method, referred to as the pilot with cyclic prefix (PCP) method, that significantly reduces PAPR compared to methods in \cite{Raviteja2019} and \cite{Zhao2020}. However, it consistently incurs higher overhead. This trade-off between PAPR, overhead, and channel estimation accuracy highlights the complexity of designing efficient channel estimation methods for OTFS systems. The authors in \cite{low} present a method with less overhead than others for integer Doppler shifts. For fractional shifts, the overhead matches that of our approach. However, its performance cannot reach the EP scheme due to its equalization in the TF domain.
In the pursuit of increasing spectral efficiency, a superimposed DD domain channel estimation method was proposed in \cite{Mishra2021}. This method assumes no pilot overhead for the system and superimposes pilots and data in each DD bin. An iterative method is used at the receiver for joint channel estimation and data detection to cancel out pilot interference. However, this method is only applicable for the doubly-under-spread channel, as it assumes that the receiver has prior knowledge of the delay and Doppler of each channel coefficient and that the covariance matrix of CSI is known at the receiver. 
 It’s important to note that all these methods utilize the DD representation of the signal for channel estimation.

This paper proposes a pilot design with a two-step channel
estimation algorithm for ZP-OTFS systems. The proposed
design addresses the disadvantages of using ZP-OTFS by 
inserting Zadoff-Chu  pilot sequences in zero bins. Such design enables us
to use the MRC detector, offering the benefits of lower computational complexity and improved SNR. The proposed method
also utilizes the time domain received sequences to estimate
the CSI which leads to reduced estimation complexity.
To combat high PAPR and reduce overhead, our proposed method spreads pilot symbols in the zero bins of ZP-OTFS system. Additionally, we use a sparse recovery algorithm to estimate CSI in two steps using time domain samples while consuming less total power compared to the method in \cite{Zhao2020}. This approach of using time domain samples for CSI estimation gives us an advantage over other estimation methods as we do not have to transform the received signal into DD domain to estimate CSI. In contrast to \cite{Mishra2021}, our proposed method estimates the delay and Doppler of each tap using a sparse recovery algorithm and assumes that the covariance matrix of CSI is unknown at the receiver. Notably, we applied the proposed method in\cite{Mishra2021} in two steps on time domain received samples instead of applying it iteratively on DD domain received samples.
Overall we propose a joint channel estimation and data detection method for the ZP-OTFS system that offers several key advantages. Notably, our proposed method does not impose high PAPR and maintains almost a constant PAPR for different numbers of subcarriers. Additionally, our method increases spectral efficiency, improves power efficiency, and utilizes the superiority of the MRC detector over other detection methods. Our method also estimates CSI using time domain samples, eliminating the need to transform the received signal into the DD domain.

The paper is structured as follows: Section II outlines the system design of the proposed method, while Section III describes the joint channel estimation and data detection process. In Section IV, simulation results of the proposed method are presented and compared with the EP method. Finally, Section V concludes the paper.

\textbf{Notation}: Boldface upper case ($\mathbf{A}$) for matrices, boldface lower case ($\mathbf{a}$) for vectors, regular lower case ($a$) for scalars, $(.)^H$ for Hermitian operator, $\text{\textbf{vec}}(\mathbf{A})$ vectorizes the matrix $\mathbf{A}$ by stacking its columns under each other, $\text{\textbf{vec}}^{-1}(\mathbf{a}) $ unvectorize the vector $\mathbf{a}$, $\otimes$ for Kronecker product, $\mathbf{F}_N$ for $N$-point  
discrete Fourier transform (DFT) matrix, $\lfloor .\rfloor$ and $\left[.\right]_M$ are floor and modulo-$M$ operator.
\section{System Design}
%In the ZP-OTFS transceiver, the DD domain divides into $M$ and $N$ bins along the delay and Doppler axis, respectively. The delay and Doppler of  the $l^{th}\left(l = 0, 1, ..., M-1\right)$ and $k^{th}\left( k = 0, 1, ..., N-1\right)$  bin in the discrete DD domain grid are $\tau=\frac{l}{M\Delta f}$ and $\nu=\frac{k}{NT_s}$, respectively. The available bandwidth is $B = M\Delta f$, where $\Delta f$ is the sub-carrier spacing. Also, $T = NT_s$ is the frame duration, where $T_s$ is the ZP-OTFS sub-symbol duration, and $T_s\Delta f=1$.

In the ZP-OTFS transceiver, the DD domain is divided into $M$ and $N$ bins along the delay and Doppler axes, respectively. The $l^{th}$ delay bin ($l = 0, 1, ..., M-1$) corresponds to a delay of $\tau=\frac{l}{M\Delta f}$, while the $k^{th}$ Doppler bin ($k = 0, 1, ..., N-1$) corresponds to a Doppler shift of $\nu=\frac{k}{NT_s}$, where $\Delta f$ is the sub-carrier spacing. The available bandwidth is $B = M\Delta f$, while $T = NT_s$ is the frame duration, where $T_s$ is the ZP-OTFS sub-symbol duration, and $T_s\Delta f=1$.

\subsection{The Transmitter}

\begin{figure}[t]
	\centering
	\begin{overpic}[width=0.45\textwidth,tics=10]{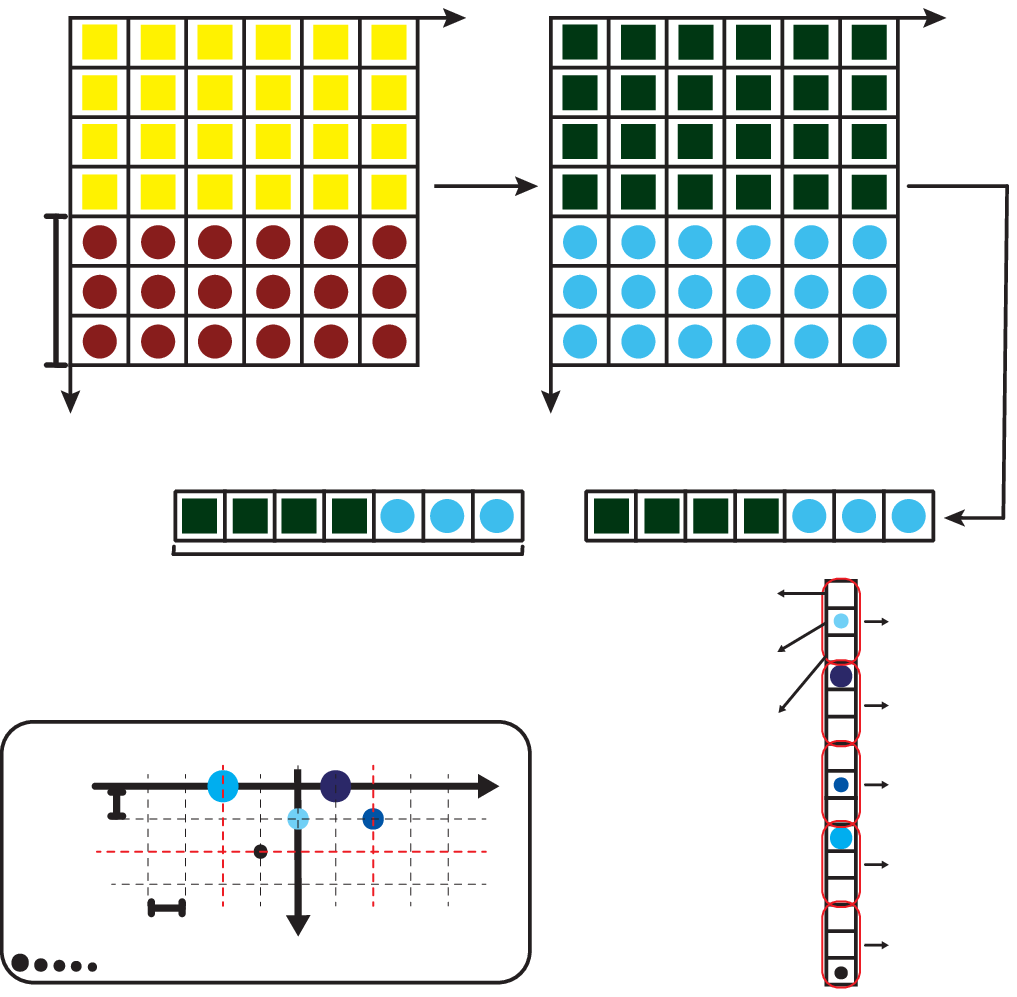}
		\put (0.3,20.8) { \scalebox{0.8}{$\frac{1}{M\Delta f}$}}
		\put (11.2,7) { \scalebox{0.8}{$\frac{1}{NT_s}$}}		
		\put (39,91) { \scalebox{0.8}{$0$}}
		\put (39,86.5) { \scalebox{0.8}{$1$}}
        \put (38.3,63) { \scalebox{0.8}{\rotatebox{0}{$M$}}}
        \put (41.8,63) { \scalebox{0.8}{\rotatebox{0}{$-$}}}        
        \put (44.4,62.8) { \scalebox{0.8}{\rotatebox{0}{$1$}}}        
        \put (83.4,91) { \scalebox{0.8}{$0$}}
        \put (83.4,86.5) { \scalebox{0.8}{$1$}}
        \put (83.1,63) { \scalebox{0.8}{\rotatebox{0}{$M$}}}
        \put (86.6,63) { \scalebox{0.8}{\rotatebox{0}{$-$}}}
        \put (89.2,63) { \scalebox{0.8}{\rotatebox{0}{$1$}}}                
		\put (-1.45,68.5) { \scalebox{0.8}{$l_{ZP}$}}
        \put (38.7,79.7) { \scalebox{0.8}{$\mathbf{XF}_N^H$}}
        \put (11,96.5) { \scalebox{0.8}{$\mathbf{X}=\mathbf{X}_d+\mathbf{X}_p$}}
		\put (56,96.5) { \scalebox{0.8}{$\mathbf{\tilde{X}}=\mathbf{\tilde{X}}_d+\mathbf{\tilde{X}}_p$}}
		\put (7.1,58.5) { \scalebox{0.8}{$0$}}
		\put (12.56,58.5) { \scalebox{0.8}{$1$}}
		\put (31.2,58.5) { \scalebox{0.8}{\rotatebox{0}{$N$}}}
		\put (34.2,58.5) { \scalebox{0.8}{\rotatebox{0}{$-$}}}
		\put (36.9,58.5) { \scalebox{0.8}{\rotatebox{0}{$1$}}}				
		\put (51.9,58.5) { \scalebox{0.8}{$0$}}
		\put (57.5,58.5) { \scalebox{0.8}{$1$}}
		\put (76,58.5) { \scalebox{0.8}{\rotatebox{0}{$N$}}}
		\put (79,58.5) { \scalebox{0.8}{\rotatebox{0}{$-$}}}
		\put (81,58.5) { \scalebox{0.8}{\rotatebox{0}{$1$}}}		
		\put (19,56) { \scalebox{0.8}{\rotatebox{0}{(a)}}}
		\put (64.5,56) { \scalebox{0.8}{\rotatebox{0}{(b)}}}
		\put (46,37) { \scalebox{0.8}{\rotatebox{0}{(c)}}}
		\put (23,1) { \scalebox{0.8}{\rotatebox{0}{(d)}}}		
		\put (74.9,1) { \scalebox{0.8}{\rotatebox{0}{(e)}}}				
		\put (3,55.5) { \scalebox{0.8}{\rotatebox{0}{$l$}}}				
		\put (41,95.6) { \scalebox{0.8}{\rotatebox{0}{$k$}}}								
		\put (48,55.5) { \scalebox{0.8}{\rotatebox{0}{$l$}}}				
		\put (86,95.6) { \scalebox{0.8}{\rotatebox{0}{$n$}}}								
		\put (68.5,22) { \scalebox{0.8}{\rotatebox{0}{$\mathbf{h}=$}}}	
		\put (-1.6,47.2) { \scalebox{0.8}{\rotatebox{0}{$\mathbf{s} = \mathbf{s}_d+\mathbf{s}_p$}}}	
		\put (47.7,46.65) { $\bigcdot$}
		\put (49.25,46.65) { $\bigcdot$}
		\put (51.1,46.65) { $\bigcdot$}
		\put (16,51) { \scalebox{0.8}{$0$}}
		\put (21,51) { \scalebox{0.8}{$1$}}
		\put (77,51) { \scalebox{0.8}{$M$}}				
		\put (80.2,51) { \scalebox{0.8}{$N$}}
		\put (40.5,51) { \scalebox{0.8}{$M$}}
		\put (44,51) { \scalebox{0.8}{$-$}}
		\put (46.5,51) { \scalebox{0.8}{$1$}}
		\put (83.2,51) { \scalebox{0.8}{$-$}}
		\put (85.9,51) { \scalebox{0.8}{$1$}}
		\put (18,41) { \scalebox{0.8}{OTFS sub-symbol}}		
%		\put (7,33.4) { \scalebox{0.8}{DD data}}				
%		\put (7,29) { \scalebox{0.8}{DD pilot}}		
%		\put (35.4,33.4) { \scalebox{0.8}{DT data}}		
%		\put (35.4,29) { \scalebox{0.8}{DT pilot}}								
		\put (20.5,25.8) { \scalebox{0.8}{$h\left(\tau,\nu\right)$}}			
		\put (25.5,5) { \scalebox{0.8}{$\tau$}}		
		\put (45.5,20) { \scalebox{0.8}{$\nu$}}				
		\put (0,8) { \scalebox{0.8}{gain}}	
		\put (57.5,40) { \scalebox{0.8}{$\tau=\frac{0}{M\Delta f}$}}			
		\put (57.5,34.3) { \scalebox{0.8}{$\tau=\frac{1}{M\Delta f}$}}
		\put (57.5,28.5) { \scalebox{0.8}{$\tau=\frac{2}{M\Delta f}$}}
		\put (81.5,38) { \scalebox{0.8}{$\nu=\frac{0}{NT_s}$}}						
		\put (81.5,29.8) { \scalebox{0.8}{$\nu=\frac{1}{NT_s}$}}
		\put (81.5,22) { \scalebox{0.8}{$\nu=\frac{2}{NT_s}$}}		
		\put (81.5,14.7) { \scalebox{0.8}{$\nu=\frac{-2}{NT_s}$}}
		\put (81.5,7) { \scalebox{0.8}{$\nu=\frac{-1}{NT_s}$}}				
		\put (82.5,79.5) { \scalebox{0.8}{vec($\mathbf{\tilde{X}}$)}}			
	\end{overpic}
	\caption{transmitter system design and channel representation: (a) the DD signal $\mathbf{X}$, (b) the DT signal $\mathbf{\tilde{X}}$, (c) the time signal $\mathbf{s}$, (d) the DD channel $h\left(\tau,\nu\right)$, (e) the channel vector $\mathbf{h}$. \includegraphics[height=\fontcharht\font`B]{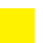}: DD data, \includegraphics[height=\fontcharht\font`B]{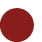}: DD pilot, \includegraphics[height=\fontcharht\font`B]{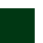}: DT data, \includegraphics[height=\fontcharht\font`B]{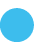}: DT pilot  }
	\label{fig:transmitted signal in delay-Doppler domain}
\end{figure}

The signal transmitted in the DD domain is depicted in Fig. \ref{fig:transmitted signal in delay-Doppler domain}a, where both information and pilot symbols are arranged. The transmitted signal in the DD domain can be represented by the matrix $\mathbf{X}=\mathbf{X}_d+\mathbf{X}_p\in \mathbb{C}^{M\times N}$, where $\mathbf{X}_d$  and $\mathbf{X}_p$ consist of information symbols and  pilot sequences, respectively.
In ZP-OTFS system, the value of $l_{ZP}$ (number of rows which must be zero in ZP-OTFS system along delay axis) must be at least equal to  $l_{max}$ , where   $l_{max}$ is the maximum  delay spread of the channel. In our approach, we consider the value  $l_{ZP}= l_{max}+1$. Consequently, the overhead of our proposed method is $(l_{max}+1)N$, which is much less than the EP scheme \cite{Raviteja2019}.    The  choice of $l_{ZP}$ ensures that, in the receiver, one last row of the received DD domain matrix becomes free of interference from data information effect. This allows us to estimate the channel in the first step by looking at the last row, which is free of data interference. Therefore, the first $M-l_{ZP}$ rows of the data matrix $\mathbf{X}_d$ carry information symbols and the others are zero, while in order to utilize all the available DD resources the last $l_{ZP}$ rows of the pilot matrix $\mathbf{X}_p$ are pilot symbols, generated from a Zadoff-Chu sequence. This arrangement ensures that the pilot sequences and information symbols in the matrix $\mathbf{X}$ do not overlap.
 Information  and pilot symbols are chosen in a way so that the average power of the transmitted signal becomes one. As a result, the average power of the signal is equal to one, given that $E\{|X[l,k]|^2\}=1$, where  $X[l,k]$ are the elements of the matrix $\mathbf{X}$.

%The time domain signal $\mathbf{s}$ is the inverse discrete Zak transform (IDZT) of $\mathbf{X}$ \cite{DDCbook},  as illustrated in fig \ref{fig:transmitted signal in delay-Doppler domain}c, as 
%\begin{equation}
%	\mathbf{s}= \mathbf{s}_d + \mathbf{s}_p = \text{IDZT}\left(\mathbf{X}\right)= \text{\textbf{vec}}\left(\mathbf{X}\mathbf{F}^H_N\right) = (\mathbf{F}_N^H\otimes\mathbf{I}_M)\mathbf{x},
%	\label{eq:vectorized_IDZT}
%\end{equation}
%where $\mathbf{s}\in\mathbb{C}^{MN\times1}$, and $\mathbf{x} = \text{\textbf{vec}}\left(\mathbf{X}\right)\in\mathbb{C}^{MN\times1}$. To prevent Inter Frame Interference (IFI), cyclic prefix (CP) is added to the beginning of the $\mathbf{s}$. The matrix $\tilde{\mathbf{X}} = \mathbf{XF}_N^H$ is defines as the DT representation of the transmitted signal. The DT signal $\mathbf{\tilde{X}}$ is shown in fig \ref{fig:transmitted signal in delay-Doppler domain}b. Besides, $\mathbf{s}_d = \text{IDZT}\left(\mathbf{X_d}\right)$ and $\mathbf{s}_p= \text{IDZT}\left(\mathbf{X_p}\right)$ are time domain data and pilot, respectively, since IDZT is a linear transformation. 

The DT representation of the transmitted signal, depicted in  Fig. \ref{fig:transmitted signal in delay-Doppler domain}b,  is defined as $\tilde{\mathbf{X}} = \mathbf{XF}_N^H$. As shown in Fig. \ref{fig:transmitted signal in delay-Doppler domain}c, the time-domain signal $\mathbf{s}$ can be obtained as the inverse discrete Zak transform (IDZT) of the DD-domain signal $\mathbf{X}$, equivalent to vectorizing the $\mathbf{\tilde{X}}$ matrix as \cite{Thaj2020}
\begin{equation}
	\mathbf{s} = \mathbf{s}_d + \mathbf{s}_p = \text{IDZT}(\mathbf{X}) = \textbf{vec}(\mathbf{XF}_N^H) = (\mathbf{F}_N^H\otimes\mathbf{I}_M)\mathbf{x},
	\label{eq:vectorized_IDZT}
\end{equation}
where $\mathbf{s} \in \mathbb{C}^{MN \times 1}$ and $\mathbf{x} = \textbf{vec}(\mathbf{X}) \in \mathbb{C}^{MN \times 1}$. Similar to the DD representation of the signal, the last $l_{ZP}$ rows along the delay axis consist of pilot sequences, since we applied the inverse DFT (IDFT) along the Doppler axis of the DD domain (along rows of the $\mathbf{X}$) to form the DT representation of the signal.
 Additionally, $\mathbf{s}_d = \text{IDZT}(\mathbf{X}_d)$ and $\mathbf{s}_p = \text{IDZT}(\mathbf{X}_p)$ represent  time-domain data and pilot vectors, respectively, since IDZT is a linear transformation.  As shown in Fig. \ref{fig:transmitted signal in delay-Doppler domain}c, pilot symbols can be distinguished from data symbols. In every $M$ samples of the time domain signal, which are defined as OTFS sub-symbols, last $l_{ZP}$ samples are pilot symbols.  Since $l_{ZP} = l_{max}+1$, in the receiver, every last sample of the OTFS sub-symbol is free of interference and is used to estimate the channel in the first step.
 To avoid inter-frame interference, a cyclic prefix (CP) is added to the beginning of $\mathbf{s}$.

\subsection{Channel Effect}

The DD domain channel representation is shown in Fig. \ref{fig:transmitted signal in delay-Doppler domain}d, along with the mathematical expression of the DD channel given by $h\left(\tau,\nu\right) = \sum_{i=1}^{Q}h_i\delta\left(\tau-\tau_i,\nu-\nu_i\right)$. Here, $\left\{h_i\right\}_{i=1}^{Q}$ represents the channel coefficients and $Q =\left(l_{max}+1\right)\left(2k_{max}+1\right)$, where $l_{max}$ and $k_{max}$ denote the maximum delay and Doppler spread of the channel, respectively, corresponding to $\tau_{max} = \frac{l_{max}}{M\Delta f}$ and $\nu_{max} = \frac{k_{max}}{NT}$. The sparsity order of the DD channel is given by the ratio of non-zero channel coefficients $P$ to $Q$. The value of $P$ represents the number of dominant reflectors, and in the receiver we estimate their delay, Doppler, and value, represented by $\tau_i$, $\nu_i$, and $h_i$, respectively \cite{9590508}.
After removing the CP, the input-output relation is described by
\begin{equation}
	\mathbf{r} = \mathbf{H}\mathbf{s} + \mathbf{w} = \mathbf{H}\mathbf{s}_d + \mathbf{H}\mathbf{s}_p + \mathbf{w}
	\label{eq:channel_effect_vector_time_domain}
\end{equation}
 where $\mathbf{w}\in\mathbb{C}^{MN\times 1}$ is the noise vector, $\mathbf{r}\in\mathbb{C}^{MN\times1}$ is the time domain received signal, and $\mathbf{H}\in\mathbb{C}^{MN\times MN}$ is the time domain channel matrix \cite{Raviteja2019a}. The channel matrix $\mathbf{H}$ is formulated as 
 \begin{equation}
 	\mathbf{H} = \sum_{i=1}^{Q}h_i\mathbf{\Pi}^{l_i}\mathbf{\Delta}^{k_i},
 	\label{eq:channel_matrix}
 \end{equation}
 where $\mathbf{\Pi}\in\mathbb{C}^{MN\times MN}$ is the standard permutation matrix, $\mathbf{\Delta }= \text{\textbf{diag}} (z^0,z^1,...,z^{MN-1})\in\mathbb{C}^{MN\times MN}$ is the diagonal Doppler matrix, and $z = e^{j2\pi/\left(MN\right)}$ \cite{DDCbook}.
By substituting  \eqref{eq:channel_matrix} into  \eqref{eq:channel_effect_vector_time_domain}, we obtain  \eqref{eq:reformulation} as
\begin{equation}
	\mathbf{r}\! \!= \!\!\left[\mathbf{\Psi}_1,  \mathbf{\Psi}_2, \cdots, \mathbf{\Psi}_Q\right]\mathbf{h} + \mathbf{w}\! \!=\! \!\mathbf{\Psi}\mathbf{h}+\mathbf{w}\! \!=\! \!\mathbf{\Psi}_d\mathbf{h}+\mathbf{\Psi}_p\mathbf{h}+\mathbf{w}, 
	\label{eq:reformulation}
\end{equation}
where $\mathbf{\Psi}_i = \mathbf{\Pi}^{l_i}\mathbf{\Delta}^{k_i}\mathbf{s}\in\mathbb{C}^{MN\times1}$, $\mathbf{h}\in\mathbb{C}^{Q\times 1}$ represents the channel vector, and $\mathbf{\Psi}= \mathbf{\Psi}d+\mathbf{\Psi}p\in\mathbb{C}^{MN\times Q}$ is the time domain signal matrix. Equation \eqref{eq:reformulation} represents the input-output relation of the channel in the time domain. 
Fig. \ref{fig:transmitted signal in delay-Doppler domain}e shows how to form $\mathbf{h}$ using $h\left(\tau,\nu\right)$, where the $i^{th}$ element of $\mathbf{h}$ corresponds to $h_i$, and its delay and Doppler are $ l_i = \left[i-1\right]_{l_{max}+1}$ and $k_i$ as
\begin{equation}
	k_i=
	\begin{cases}
		\lfloor \frac{i-1}{l_{max}+1}\rfloor, & \!\!\!\!\text{if}\!\  \lfloor \frac{i-1}{l_{max}+1}\rfloor\!<=k_{max} ,\\
		\lfloor \frac{i-1}{l_{max}+1}\rfloor-(2k_{max}+1), &\!\!\!\!\text{if}\ \! \lfloor \frac{i-1}{l_{max}+1}\rfloor\!>k_{max}.
	\end{cases}
	\label{eq:k_i}
\end{equation}

In \cite{Mishra2021}, the authors proposed a reformulation in  \eqref{eq:reformulation} in the DD domain. They assumed that the vector $\mathbf{h}$ is not sparse and considered that the delay and Doppler of each tap is known in the receiver, allowing it to only estimate the  channel gain $h_i$. However, their method was not able to estimate the taps' delay and Doppler.
Our proposed reformulation in  (4) allows us to estimate the CSI using time domain sequences, eliminating the need to transform the received signal into the DD domain. Additionally, our approach takes into account the $\mathbf{h}$'s sparsity, so we do not need to know the taps' delay and Doppler.
 Finally, the paper proposes estimating the channel vector $\mathbf{h}$ using the orthogonal matching pursuit (OMP) algorithm, which can be applied to  \eqref{eq:reformulation}. In our research we have utilized OMP algorithm since its low complexity makes it computationally efficient, while its good performance ensures accurate results \cite{8577023}. 
\subsection{Receiver}
%As shown in fig \ref{fig:receiver}a and fig \ref{fig:receiver}b, the DT representation of the received signal is $\mathbf{\tilde{Y}}=\text{vec}^{-1}\left(\mathbf{r}\right)$, where $\mathbf{r}=\mathbf{r}_d+\mathbf{r}_d$. It means that the received signal is the superposition of the channel effected data and pilot. Then using $N$-point  DFT along the time axis, we have the DD domain representation of the received signal $\mathbf{Y}=\mathbf{\tilde{Y}}\mathbf{F}_N$ \cite{Thaj2020}. In this paper, we estimate $\mathbf{h}$, using $\mathbf{s}$, and then we fed the DT signal into the MRC detector.
%
%As depicted in Fig. \ref{fig:receiver}a and Fig. \ref{fig:receiver}b, the received signal can be represented in the time domain as $\mathbf{r}=\mathbf{r}_d+\mathbf{r}_p$, where $\mathbf{r}_d$ and $\mathbf{r}_p$ are the channel-effected data and pilot signals, respectively. To obtain the corresponding DD domain representation, we apply an $N$-point Discrete Fourier Transform (DFT) to $\mathbf{\tilde{Y}}$ along the time axis, yielding $\mathbf{Y}=\mathbf{\tilde{Y}}\mathbf{F}_N$, where $\mathbf{\tilde{Y}}=\text{vec}^{-1}\left(\mathbf{r}\right)$ is the DT representation of the received signal. This approach has been previously described in \cite{Thaj2020}.

\begin{figure}
  \centering
	\begin{overpic}[width=0.45\textwidth,tics=10]{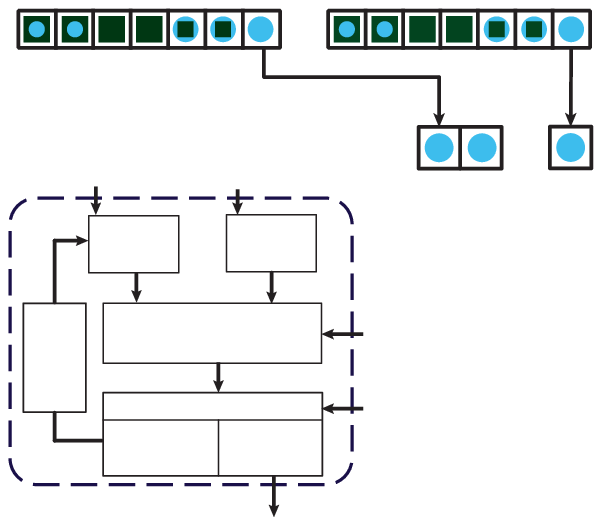}
					\put (12,48) { \scalebox{0.8}{$\mathbf{r}$ (2nd step)}}
					\put (33,48) { \scalebox{0.8}{$\mathbf{r'}_p$(1st step)}}
					\put (56,25) { \scalebox{0.9}{$\mathbf{r}$}}		
					\put (17,39.5) { \scalebox{0.8}{OMP on}}										\put (36.5,39.5) { \scalebox{0.8}{OMP on}}					
					\put (18.3,35.9) { \scalebox{0.8}{Eq. \eqref{eq:iterative_received_signal_sparse_recovery}}}					
%					\put(58.5,32.5){\scalebox{0.8}{$\mathbf{\hat{\Psi}}\mathbf{h}$}}									
%%					\put(56,32.5){\scalebox{0.8}{$=$}}														
%%					\put (74.4,32.5) { \scalebox{0.8}{$\mathbf{r'}_p$}}					
					\put(39,35.9){\scalebox{0.8}{Eq. \eqref{eq:time_coarse_estimation_formula}}}									
%					\put(79.4,32.5){\scalebox{0.8}{$=$}}																			
					\put(22,25.2){\scalebox{0.8}{Apply Eq. \eqref{eq:data_without_pilot_time_receiver} on $\mathbf{r}$}}																													
%					
%%					\put(62,22.6){\scalebox{0.8}{Cancellation}}
							\put(26,14.7){\scalebox{0.8}{MRC detector}}
					\put(45,31.1){\scalebox{0.8}{$\mathbf{\hat{h}}$}}
					\put(25.9,31.1){\scalebox{0.8}{$\mathbf{\hat{h}}$}}
										\put(57.3,15){\scalebox{0.8}{$\mathbf{\hat{h}}$}}
	\put(38,19){\scalebox{0.8}{$\mathbf{r}_d$}}				
							\put(41,11){\scalebox{0.8}{DD}}
										\put(39,8){\scalebox{0.8}{output}}				
					\put(26,11){\scalebox{0.8}{DT}}
		\put(24,8){\scalebox{0.8}{output}}									
		\put(10,16){\rotatebox{90}{\scalebox{0.8}{$\textbf{vec}\left(\mathbf{\hat{\tilde{X}}}_d\right)$}}}											
\put (13.5,6) { \scalebox{0.7}{\rotatebox{0}{$\mathbf{\hat{\tilde{X}}}_d$}}}		
\put (44,1) { \scalebox{0.7}{\rotatebox{0}{$\mathbf{\hat{X}}_d$}}}		
	\put (65.8,46.4) { \scalebox{0.7}{$0$}}
\put (72,46.4) { \scalebox{0.7}{$1$}}
\put (82,46.4) { \scalebox{0.7}{\rotatebox{0}{$N$}}}
\put (84.4,46.4) { \scalebox{0.7}{\rotatebox{0}{$-$}}}
\put (86.7,46.4) { \scalebox{0.7}{\rotatebox{0}{$1$}}}

	\put (8.5,73) { \scalebox{0.7}{$0$}}
\put (16,73) { \scalebox{0.7}{$1$}}
\put (37.5,73) { \scalebox{0.7}{\rotatebox{0}{$M$}}}
\put (40.5,73) { \scalebox{0.7}{\rotatebox{0}{$-$}}}
\put (43,73) { \scalebox{0.7}{\rotatebox{0}{$1$}}}

\put (55,51.5) { \scalebox{0.7}{\rotatebox{0}{$\mathbf{r'}_p=$}}}

\put (0,68.5) { \scalebox{0.7}{\rotatebox{0}{$\mathbf{r}=$}}}

\put (80,73) { \scalebox{0.7}{\rotatebox{0}{$M$}}}
\put (83,73) { \scalebox{0.7}{\rotatebox{0}{$N$}}}
\put (85.5,73) { \scalebox{0.7}{\rotatebox{0}{$-$}}}
\put (87.7,73) { \scalebox{0.7}{\rotatebox{0}{$1$}}}
%\put (58,81.5) { \scalebox{0.7}{\rotatebox{0}{$\mathbf{\tilde{Y}}=\mathbf{\tilde{Y}}_d+\mathbf{\tilde{Y}}_p$}}}				
%\put (17,46.1) { \scalebox{0.7}{\rotatebox{0}{$\mathbf{\hat{X}}_d$}}}
%\put (76.2,0.65) { \scalebox{0.7}{\rotatebox{0}{$\mathbf{\hat{X}}_d$}}}
		\put (44.7,68.3) { $\bigcdot$}
\put (46.3,68.3) { $\bigcdot$}
\put (47.9,68.3) { $\bigcdot$}
		\put (76,51.3) { $\bigcdot$}
\put (77.6,51.3) { $\bigcdot$}
\put (79.2,51.3) { $\bigcdot$}
\put(93,68.5){\scalebox{0.8}{(a)}}		
\put(93,51.5){\scalebox{0.8}{(b)}}		
\put(30,0){\scalebox{0.8}{(c)}}		
\put(57,25){\scalebox{1}{\rotatebox{180}{{\rotatebox{0}{$\begin{cases}
						 & \\
						 & \\
						 & \\
						 & \\
						 & \\
						 & \\
						 & 
					\end{cases}   $}}}}}		
\put (67,25) { \scalebox{0.7}{Joint channel estimation}}												
\put (64,21) { \scalebox{0.7}{and data detection algorithm}}												
%																\put(20,9){\scalebox{0.8}{(d)}}		
%																				\put(65,0.5){\scalebox{0.8}{(e)}}		
%			\put(69.5,37.4){\scalebox{0.8}{$\mathbf{\hat{h}}$}}									
						
	\end{overpic}
  \caption{ receiver system design: (a) the time vector $\mathbf{r}$, (b) channel-affected pilot vector without data interference $\mathbf{r'}_p$ , (c) proposed algorithm.  \includegraphics[height=\fontcharht\font`B]{fig/dtpilot.eps}: channel-affected pilot symbols without data interference \includegraphics[height=\fontcharht\font`B]{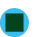}: channel-affected pilot symbols with data interference, \includegraphics[height=\fontcharht\font`B]{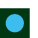}: channel-affected data symbols with pilot interference }\label{fig:receiver}
\end{figure}

As depicted in Fig. \ref{fig:receiver}a, the received signal can be represented in the time domain as $\mathbf{r}=\mathbf{r}_d+\mathbf{r}_p$, where $\mathbf{r}_d$ and $\mathbf{r}_p$ are the channel-effected data and  pilot, respectively. To obtain the corresponding DD domain representation of the time domain signal, we apply an $N$-point DFT to $\mathbf{\tilde{Y}}$ along the time axis, yielding $\mathbf{Y}=\mathbf{\tilde{Y}}\mathbf{F}_N$, where $\mathbf{\tilde{Y}}=\text{vec}^{-1}\left(\mathbf{r}\right)$ is the DT representation of the received signal. This approach is known as the discrete Zak transform (DZT).

In the receiver, we use time domain channel-affected pilot symbols in $\mathbf{r}$ to estimate the channel vector $\mathbf{h}$, rather than its DD representation. We then employ an MRC detector to detect the information symbols, thanks to our pilot design which enables the use of an MRC detector in the receiver. To ensure accurate channel estimation, we use the entire time domain received signal in the next step, comprising both channel-affected data and pilot, to estimate the channel vector.

\section{Joint Channel Estimation and Data Detection}
%In this paper, two-step channel estimation and data
%detection is proposed. In the first step, the channel is
%estimated coarsely. Afterward, the detection is employed
%using MRC detector after interference cancellation. In the
%second step, the detected data and pilot sequences are
%used jointly as pilot to estimate channel accurately.
This paper uses a two-step method for channel estimation and data detection. The first step involves obtaining a preliminary estimate of the channel, followed by interference cancellation and the implementation of the MRC detector for data detection. In the second step, both the detected data and pilot sequences are jointly used as pilots to refine the channel estimation, resulting in a more accurate channel representation.

Our  method builds upon the approach  in \cite{Mishra2021}, but with several key differences. Our pilot design allows for a more accurate  channel estimation in the first step, resulting in improved performance compared to \cite{Mishra2021}. While the method in \cite{Mishra2021} requires multiple iterations to accurately estimate the channel, our approach achieves a high level of accuracy in just two steps. Additionally, our pilot design enables the use of a MRC detector, which detects data in the DT domain. Since we estimate the channel using time-domain sequences, there is no need to transform the received data into the DD domain.
%The proposed channel estimation algorithm uses the time domain
%representation of the received signal. By applying a sparse recovery algorithm,
%the sparse nature of the DD representation of the wireless channel can be
%exploited.
%OMP algorithm is used here to estimate the channel vector h. As our estimation
%method uses less number of pilots, the estimation accuracy can not reach the embedded pilot scheme, but is acceptable. To reduce estimation error and
%iterative interference cancellation is used here.
%The proposed channel estimation method is based on sparse recovery algorithms. Here the algorithm that is used to estimate channel vector is OMP. Also, our proposed method consist of two step. In the first step a coarse channel estimation is done and in the next step an iterative algorithm and interference cancellation is used to improve estimation error.
\subsection{First Step}
%Considering the entries in $\mathbf{r}$ without data interference, we can rewrite \eqref{eq:reformulation}  as
% To estimate channel vector $\mathbf{\hat{h}}$, we perform OMP on \eqref{eq:time_coarse_estimation_formula}.
%To detect data, we cancel out pilot effect of the received signal as
%\begin{equation}
%	\mathbf{r}_d = \mathbf{r} - \mathbf{\Psi}_p\mathbf{\hat{h}} = \mathbf{\Psi}_d\mathbf{h} + \mathbf{\Psi}_p\left(\mathbf{h}-\mathbf{\hat{h}}\right) + \mathbf{w}  ,
%	\label{eq:data_without_pilot_time_receiver}
%\end{equation}
%where $\mathbf{r}_d\in\mathbb{C}^{MN\times 1}$ is the received data vector without pilot effect. Afterward, $\mathbf{\hat{h}}$ and $\mathbf{r}_d$ are fed to the MRC detector to the detect data vector $\mathbf{\hat{x}}_d$. Also, $\mathbf{\hat{s}}_d$ and $\mathbf{\hat{\Psi}}_d$ are  the time domain vector and the matrix of the detected data, respectively. The term $\mathbf{\Psi}_p\left(\mathbf{h}-\mathbf{\hat{h}}\right)$ in \eqref{eq:data_without_pilot_time_receiver} is the interference remained  due to the channel estimation error. To reduce such interference, the second step  is needed to estimate channel vector with good performance. Consequently, the detection error is reduced.
%%

We estimate the CSI using channel-affected pilot entries in $\mathbf{r}$ that are free from data interference. These entries form a new vector $\mathbf{r'}_p \in\mathbb{C}^{N\times 1}$, shown in Fig. \ref{fig:receiver}b. We  rewrite \eqref{eq:reformulation} as
\begin{equation}
	\mathbf{r'}_p = \mathbf{\Psi'}_p\mathbf{h} + \mathbf{w'},
	\label{eq:time_coarse_estimation_formula}
\end{equation} 
 where $\mathbf{w}'\in\mathbb{C}^{N\times 1}$ represents the noise vector for $\mathbf{r'}_p$ entries, and $\mathbf{\Psi'}_p\in\mathbb{C}^{N\times Q}$ is the time-domain pilot matrix of the first step. It is important to note that the matrix $\mathbf{\Psi'}_p$ only consists of the rows of $\mathbf{\Psi}$ that contribute to forming $\mathbf{r'}_p$.  
Now, performing OMP on \eqref{eq:time_coarse_estimation_formula}  yields the estimated channel vector $\mathbf{\hat{h}}$ \cite{8577023}.
 Afterward, to detect data, we cancel out the channel-affected pilot symbols of the received signal $\mathbf{r}$  
\begin{equation}
	\mathbf{r}_d = \mathbf{r} - \mathbf{\Psi}_p\mathbf{\hat{h}} = \mathbf{\Psi}_d\mathbf{h} + \mathbf{\Psi}_p\left(\mathbf{h}-\mathbf{\hat{h}}\right) + \mathbf{w},
	\label{eq:data_without_pilot_time_receiver}
\end{equation}
where  $\mathbf{r}_d\in\mathbb{C}^{MN\times 1}$ denotes the received channel-affected data vector with the influence of channel-affected pilot 	symbols eliminated. In our proposed pilot scheme, the last entries of each OTFS sub-symbol are initially occupied by pilot symbols. However, these pilot symbols are subsequently eliminated. Following their removal, we employ the MRC detector to detect the information symbols from the vector $\mathbf{r}_d$. This approach allows us to effectively utilize the MRC detector in the presence of pilot symbols.  At this stage, the DT output representation of the detected data is denoted by $\mathbf{\hat{\tilde{X}}}_d$. Since the MRC detector produces DT output, we utilize this DT output of the detected symbol in the next step, as illustrated in Fig. \ref{fig:receiver}c.
The term $\mathbf{\Psi}_p\left(\mathbf{h}-\mathbf{\hat{h}}\right)$ in \eqref{eq:data_without_pilot_time_receiver} indicates the amount of interference that remains due to the imperfect channel estimation. To minimize this interference, we need a second step to obtain a more accurate estimate of the channel vector, which will improve the detection performance. The first step and the second step of our proposed method are shown in Fig. \ref{fig:receiver}c. This figure indicates that we use $\mathbf{r'}_p$ in the first step to estimate the channel vector $\mathbf{\hat{h}}$ and after removing the channel-affected pilot symbols, the MRC detector uses $\mathbf{\hat{h}}$ and $\mathbf{r}_d$ to detect  information symbols named $\mathbf{\hat{\tilde{X}}}_d$ in the DT domain.
\subsection{Second Step}
In this step, the time domain pilot sequences $\mathbf{s}_p$ and detected data $\mathbf{\hat{s}}_d =\textbf{vec}\left(\mathbf{\hat{\tilde{X}}}_d\right)$ from the first step are jointly used as pilot to accurately estimate the channel and reduce detection errors, shown in Fig. \ref{fig:receiver}c. To achieve this, we rewrite \eqref{eq:reformulation} as
\begin{equation}
	\mathbf{r}\! =\! \left(\!\mathbf{\Psi}_p\!+\mathbf{\hat{\Psi}}_d\!\right)\mathbf{h}\!+\!\left(\!\mathbf{\Psi}_d\!-\!\mathbf{\hat{\Psi}}_d\!\right)\mathbf{h}\!+\!\mathbf{\hat{w}} \!=\! \left(\!\mathbf{\Psi}_p\!+\!\mathbf{\hat{\Psi}}_d\!\right)\mathbf{h}\!+\!\mathbf{\bar{w}},
	\label{eq:iterative_received_signal_sparse_recovery}
\end{equation}
where the term $\mathbf{\hat{\Psi}}=\mathbf{\Psi}_p+\mathbf{\hat{\Psi}}_d$ is considered as the joint pilot and $\mathbf{\hat{\Psi}}_d$ are time domain detected data matrix that is formed using $\mathbf{\hat{s}}_d$ according to \eqref{eq:reformulation}. Unlike the first step, this approach uses all the entries of $\mathbf{r}$ to estimate the channel, including those with data interference. The channel vector is then estimated using the OMP algorithm on \eqref{eq:iterative_received_signal_sparse_recovery}.

%To detect data, pilot contamination into the data is
%removed according to \eqref{eq:data_without_pilot_time_receiver} and data is detected using MRC
%detector. The detected data is named $\mathbf{\hat{X}}_d$ and is shown
%in fig 2d.  Now, as the channel vector is estimated with
%lower error than the first step, the MRC detector has
%better performance. The proposed algorithm is illustrated
%as a joint channel estimation and data detection in Figure \ref{fig:receiver}(e).

As shown in Fig. \ref{fig:receiver}c, to detect data, similar to previous step, we remove the channel-affected pilot from the received signal using \eqref{eq:data_without_pilot_time_receiver}, and employ the MRC detector to detect the DD representation of the data matrix $\mathbf{\hat{X}}_d$. As the channel vector is estimated more accurately than in the first step, the MRC detector results in better performance in the second step. 
\section{Simulation Results}
In this section, we evaluate the performance of our proposed channel estimation method. The system operates at a carrier frequency of $f_c=4$ GHz and a subcarrier spacing of $\Delta f=15$ kHz, with a maximum speed of the fading channel set to $v=500$ km/h. Our method demonstrates a  contribution by achieving lower overhead and PAPR compared to the EP scheme, and lower overhead with similar PAPR compared to the PCP scheme. This competitive performance validates the effectiveness of our method and its potential for practical implementation in future wireless communication systems. We use the EVA channel model and Jakes' formula to generate DD channel parameters \cite{DDCbook}, assuming integer delay and Doppler for each tap. Our pilot design could potentially be extended to work with fractional channel parameters in future researches. The value of single pilot tone in the EP scheme is $\sqrt{(4k_{\text{max}}+1)(2l_{\text{max}}+1)}$, while our method considers less pilot overhead, resulting in lower dedicated pilot power. Parameters $M$ and $N$ are set to 64 for simulating BER and normalized mean square error (NMSE) curves while 4-QAM alphabet is used.

\begin{figure}
	\centering
	\vspace{-0.25cm}
	\subfloat[\label{2a}]{%
		\includegraphics[width=0.48	\linewidth]{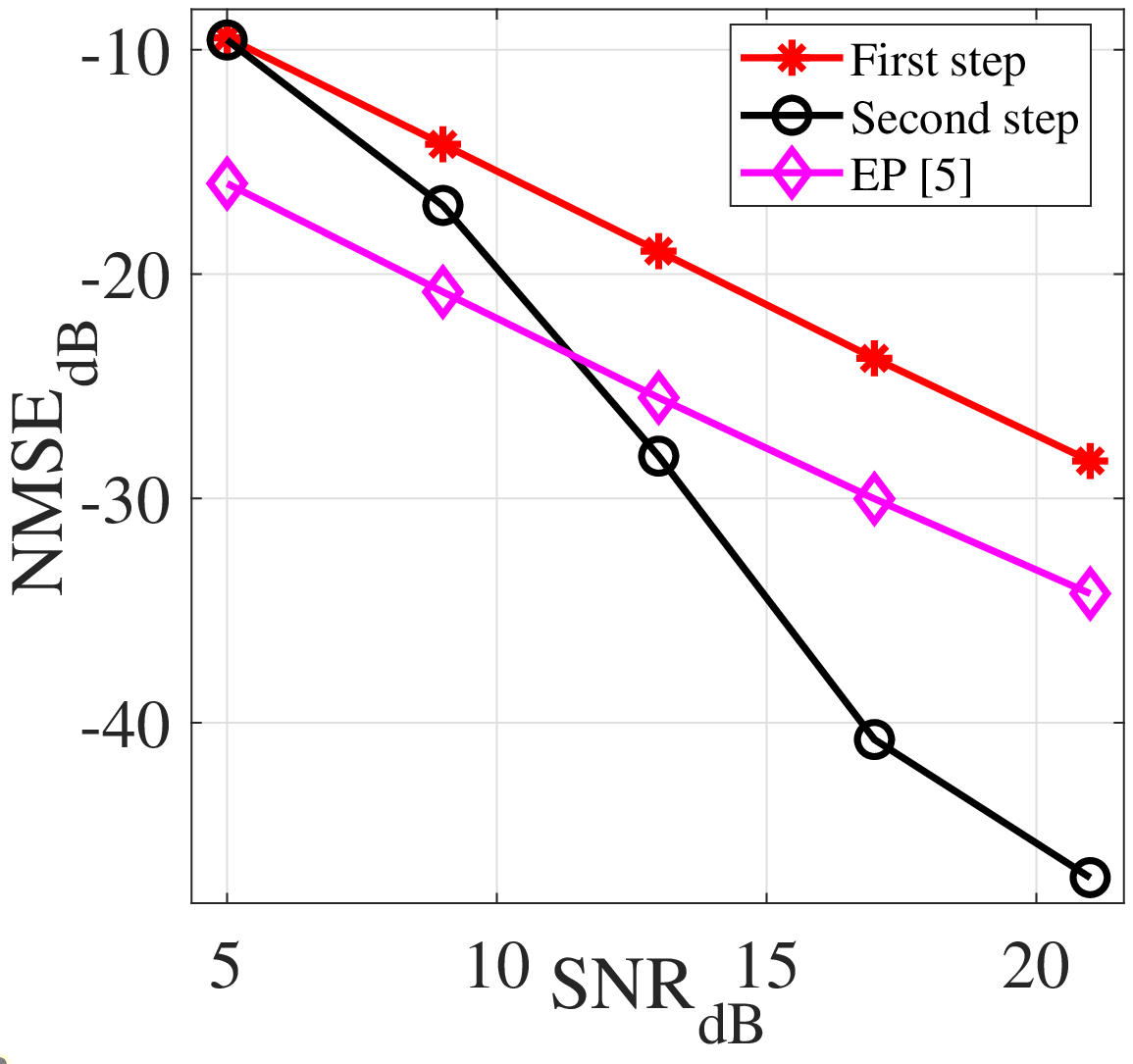}}
%	\hfill
		\hspace{-1em}
	\subfloat[\label{2b}]{%
		\includegraphics[width=0.48\linewidth]{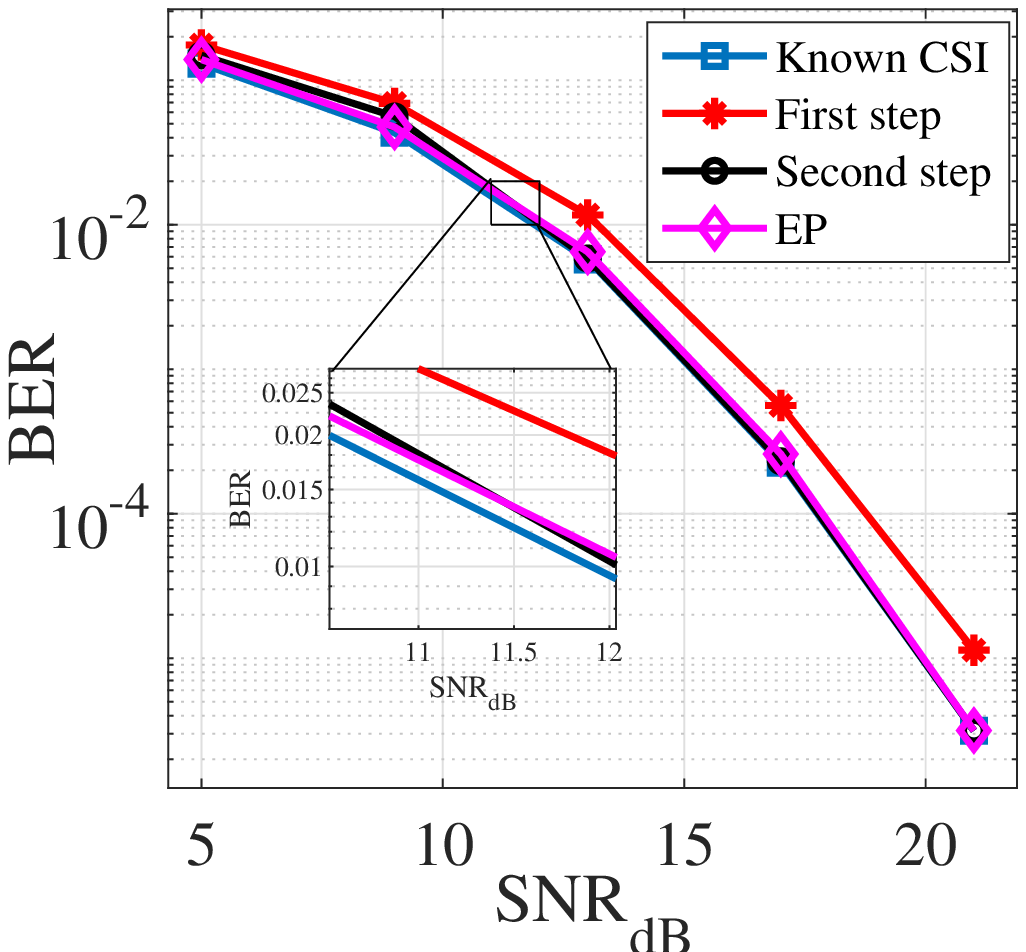}}
	\caption{(a) The NMSE of the proposed method (first step and second step) with EP, (b) The BER  of the proposed method with EP and known CSI.}
 \vspace{-0.5cm}
	\label{fig2}
\end{figure}

%Fig \ref{2a} illustrates  NMSE versus SNR for proposed estimation method and EP. This figure shows that our proposed
%method in the first step cannot reach to EP performance.
%However, in the second step of the proposed method,
%estimation error is much lower than that of EP for high
%SNRs. It is clear that in the low SNRs, the NMSE of the
%second step is lower than EP. This is due to the fact that
%the channel estimation in the first step is poor in the low
%SNRs. Consequently, the first step data detection is not
%good enough to be used in the second step.
Fig. \ref{2a} compares the NMSE versus SNR for our method and the EP scheme. The first step of our method underperforms EP, but the second step outperforms EP at high SNRs due to less error in the first step. However, it underperforms at low SNRs due to suboptimal channel estimation. A crossover point occurs at an SNR of about 11 dB, indicating a  performance transition . This crossover will impact the BER curve.

Fig. \ref{2b} displays the BER versus SNR, with the second step’s BER matching the EP and known CSI. Despite the first step’s NMSE not reaching EP, its BER is comparable. The crossover in the BER curves of the second step and EP is due to the crossover in NMSE of channel estimation, emphasizing the impact of channel estimation quality on BER.

\begin{figure}
	\centering
	\subfloat[\label{3a}]{%
		\includegraphics[width=0.515\linewidth]{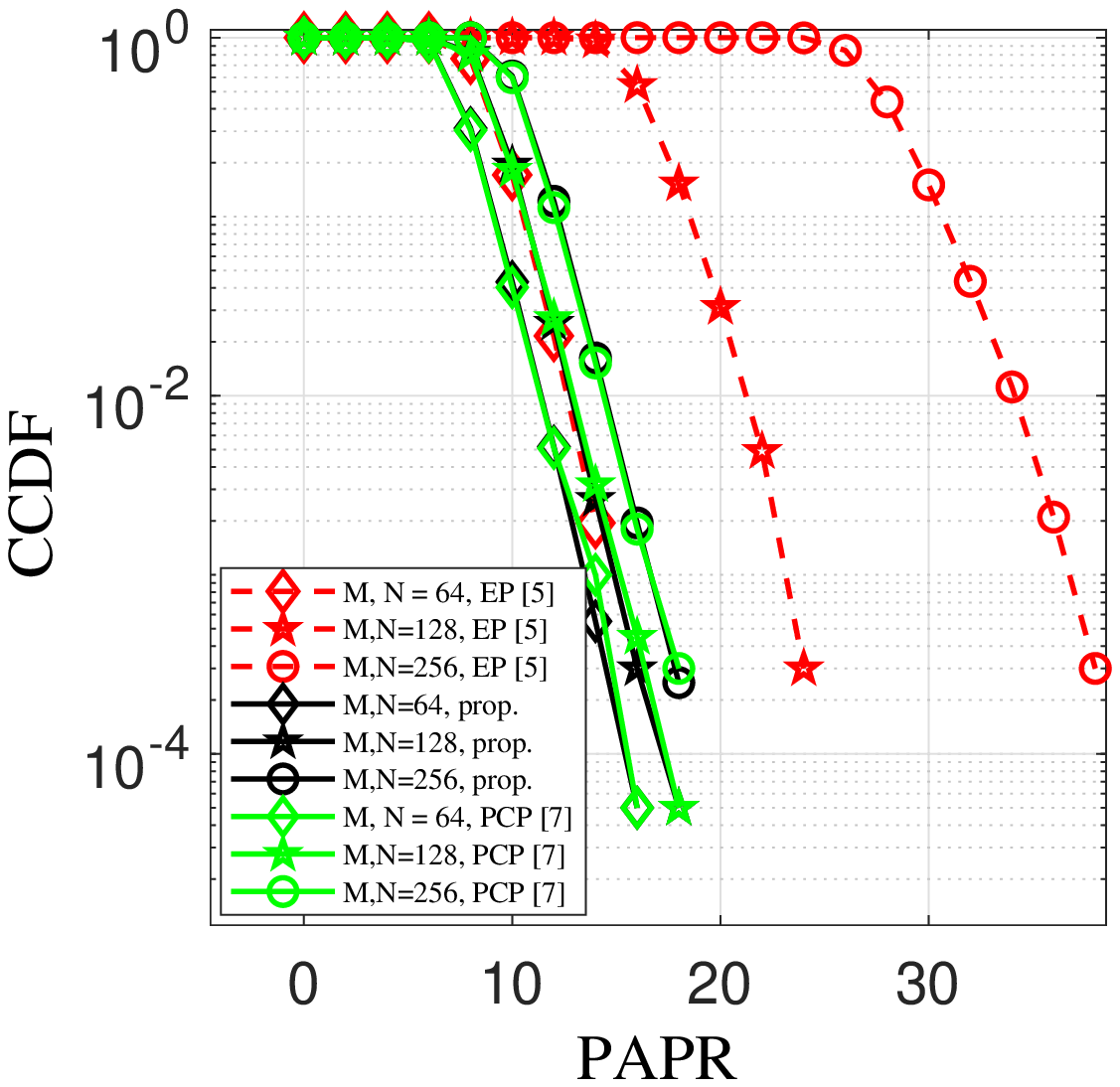}}
		\hspace{-2.3em}
	\subfloat[\label{3b}]{%
		\includegraphics[width=0.56\linewidth]{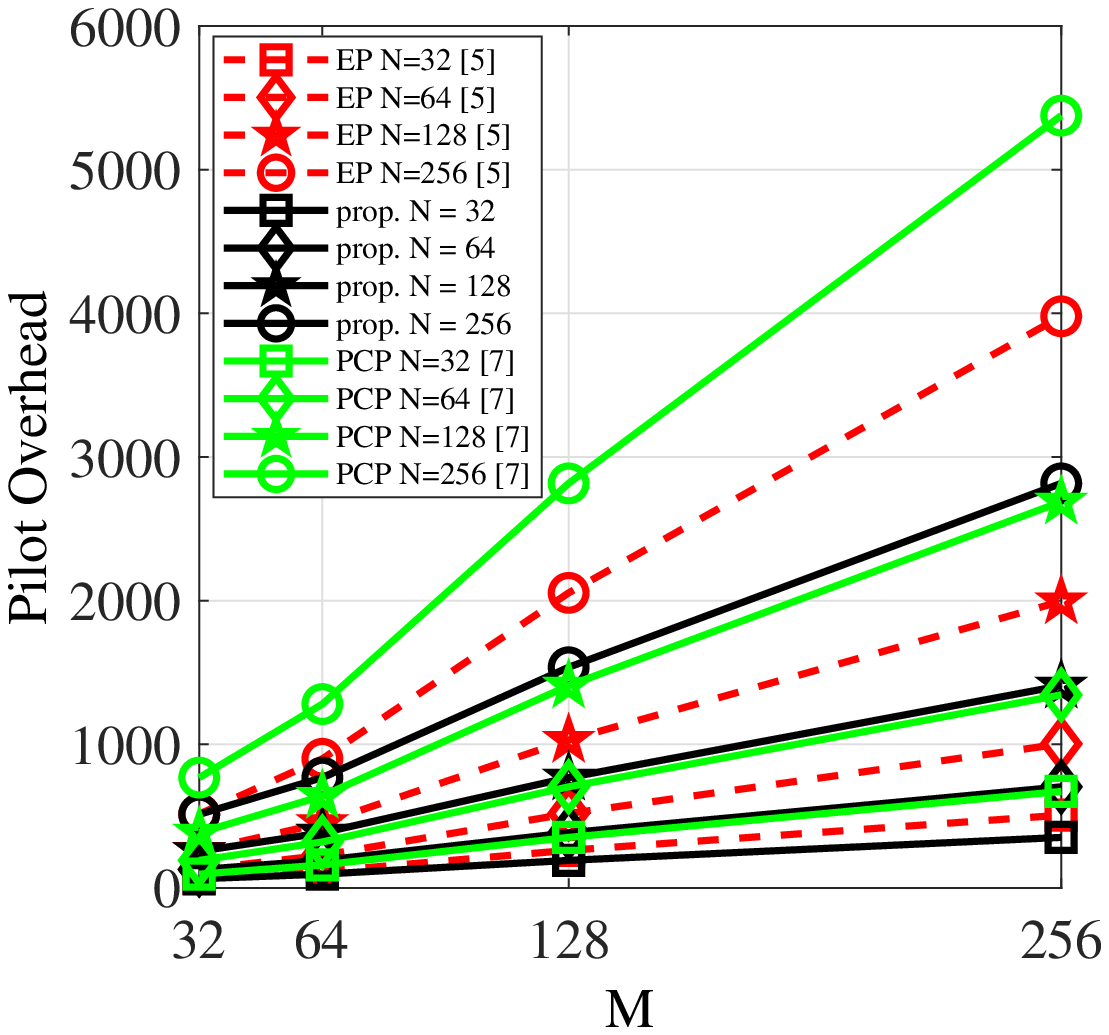}}
	\caption{(a) CCDF versus PAPR of the transmitted signal, (b) Number of pilot overhead versus $M$ for different $N$.}
	\label{fig3}
\end{figure}

Fig. \ref{3a} illustrates the complementary cumulative distribution function (CCDF) for the PAPR of the transmitted signal for different values of $M$ and $N$. The PAPR of the proposed method is significantly lower than that of the EP signal. This can be attributed to the fact that in the EP scheme, only one high power single pulse in the DD domain is utilized. Consequently, as $M$ and $N$ increase, leading to an increase in the channel parameters $l_{max}$ and $k_{max}$ respectively, more power is allocated to a single pilot in the EP scheme, resulting in a high PAPR. In contrast, our proposed system distributes pilots across the DD domain, which results in a low PAPR. Interestingly, this low PAPR changes only slightly with an increase in $M$ and $N$. It is also worth noting that  the PCP method exhibits a similar PAPR to our proposed system.

Fig. \ref{3b} illustrates the number of overhead bins for the proposed method, the PCP method, and the EP scheme for different values of $M$ and $N$. As $N$ increases, the overhead of the system also increases. In the EP and PCP methods, this is due to the additional guard used to prevent interference from data symbols to the pilot symbols, which results in increased overhead. However, our proposed method circumvents this issue by estimating the channel in two steps. This approach not only decreases the overhead, making it more efficient than both the EP scheme and the PCP method, but also makes our proposed method particularly effective for systems where low overhead is crucial.

%
%By increasing $N$, pilot overhead increases
%dramatically for EP scheme compared to the proposed
%method. This is due to the fact that the EP overhead
%depends on  $k_{max}$,
%while the proposed method does not depend on $k_{max}$.

%\begin{figure}[!t]
%	\centering
%	\includegraphics[width=1\linewidth]{fig/papr.eps}
%	\caption{CCDF versus PAPR for the transmitted signal.}
%	\label{fig_3}
%\end{figure}

\section{conclusion}
In this paper, a two-step joint channel estimation and data detection scheme for ZP-OTFS systems based on sparse recovery algorithms was proposed. In the first step, the OMP algorithm was used for channel estimation, and in the second step, data detection was performed using the MRC detector. The proposed method offered significant improvements in both spectral and power efficiency while imposing a low PAPR on the transmitter without affecting the system performance.
\bibliography{main}
\bibliographystyle{IEEEtran}

\vfill

\end{document}